\documentclass[english]{article}
\usepackage[T1]{fontenc}
\usepackage{amsmath}
\usepackage{graphicx}
\usepackage{babel}
\begin{document}
\global\long\def\ket#1{\left|#1\right\rangle }%

\global\long\def\bra#1{\left\langle #1\right|}%

\global\long\def\braket#1#2{\left\langle #1\left|#2\right.\right\rangle }%

\global\long\def\ketbra#1#2{\left|#1\right\rangle \left\langle #2\right|}%

\global\long\def\braOket#1#2#3{\left\langle #1\left|#2\right|#3\right\rangle }%

\global\long\def\mc#1{\mathcal{#1}}%

\global\long\def\nrm#1{\left\Vert #1\right\Vert }%

\title{Methods for measuring noise, purity changes, and entanglement entropy
in quantum devices and systems}
\author{Raam Uzdin}
\maketitle
\begin{abstract}
We present methods for evaluating the rate of change in quantities
during quantum evolution due to coupling to the environment (dissipation
hereafter). The protocol is based on repeating a given quantum circuit
(or quantum operation) twice, thrice, and so on, and measuring an
expectation value after each number of repetitions. We start by applying
this method for measuring the rate of purity changes in quantum circuits.
This provides direct information on the quality of the circuit. Furthermore,
the presented scheme enables to distill the dissipative contribution
in the changes of quantities such as energies and coherence. In particular,
this can be applied to the local Hamiltonians of specific qubits.
Thus, our approach can be used to locate ``hotspots'' where the
dissipation takes place. A variant of this method can be used to measure
the entanglement buildup in quantum circuits. These methods are scalable
as they involve only a few observables which are relatively easy to
measure in NISQ devices.
\end{abstract}

\section{Introduction}

The premise of quantum technology and quantum computing is to provide
a dramatic improvement compared to classical devices. In particular,
quantum computers and simulators should help solve with finite resources
(time, memory, energy, etc.) problems that would otherwise require
unrealistic resources. The complicated nature of quantum evolution
makes error detection and diagnostics very challenging. If the output
cannot be computed by some other means, it is difficult to crosscheck
and rule out the possibility that the device is either defective to
begin with, or it had malfunctioned during its operation. This is
especially true in the several dozen qubits NISQ (noisy intermediate
scale quantum) devices that are available today. More importantly,
as explained next, errors can arise either due to calibration imperfections
or due to external noise (environment). Operationally, it is important
to distinguish between the two types of errors. Unfortunately, this
task gets more difficult as the circuits get larger.

Although our results are not restricted to quantum computers and simulators
it is instructive to have in mind a device such as a quantum processor
with multiple interacting qubits. We start by describing several challenges
associated with quantum diagnostics. What connects these topics is
that we can address them with our method. In the sections that follow,
we apply our methods to these problems. 

The methods described in this paper are registered as US provisional
patent 63/260501

\subsection{Source of errors}

There are several sources of errors in NISQ devices and the two main
ones are:
\begin{itemize}
\item Coherent errors: the device is well isolated from the environment,
and the evolution is unitary. However, the device is not executing
the unitary operation (``the circuit'' in quantum computers) it
was instructed to run.
\item Incoherent error: interaction with some known or unknown environment
leads to non-unitary evolution. This can be either non-unital maps
such as thermalization (e.g. spontaneous emission) or unital maps
such as decoherence or depolarizing channels (in a unital map, the
fully mixed state is a fixed point of the map). 
\end{itemize}
Other sources of errors include state preparation errors and readout
errors. However, the first is typically very small and the latter
can be resolved by detector calibration procedures.

It is of prime importance to distinguish between coherent and incoherent
errors. Coherent errors occur because some parameters in the circuits
are not optimally calibrated. In principle, coherent errors can always
be fixed by another unitary transformation in the Hilbert space of
the original circuit. In contrast, incoherent errors, e.g. decoherence,
spontaneous emission, and depolarizing channels, cannot be removed
by unitary operations on the circuit alone. Resolving coherent errors
from incoherent errors can guide developers and experimentalists where
to focus their efforts and also validate if their efforts successfully
mitigated the error. 

\subsection{Holistic vs. One-Circuit Diagnostics}

Holistic methods such as quantum volume, randomized benchmarking,
and cross entropy benchmarking , characterize the device as a whole.
Holistic scores aim to assure a certain level of performance for any
circuit. This is appealing for quantum computers where various algorithms
may be executed on the same machine. In one-circuit diagnostics, the
circuit that executes the computation is given and its performance
on the existing hardware is evaluated. Although it may seem that the
holistic approach is more useful, the one-circuit approach has its
own merits, and in many cases it will be the first choice in quantifying
performance.

1. A given hardware may execute some circuits with sufficiently good
fidelity while in others the fidelity is quite poor. It could be that
although the holistic score is very low (a poor device), a clever
choice of qubits and gates (implementation map) may lead to good fidelity.
This is especially relevant for the presently available NISQ devices.
After making this choice there is no point in recalculating the holistic
score of the whole device (it will remain the same). Instead, the
fidelity in the specific choice of implementation map should be evaluated
directly. Although it is possible to apply holistic methods to evaluate
the selected implementation map the obtained score is not holistic
anymore and it may consume a lot of resources compared to other methods
for evaluating a specific circuit.

2. In the opposite scenario there is a reasonably good holistic score,
but for the circuit of interest, the fidelity is poor. This could
be due to an unusually large usage of a noisy gate. A one-circuit
diagnostic scheme can be significantly better at detecting such problems.

3. In developing the hardware of a quantum computer, there is often
a known gate that is susceptible to noise that the developers want
to minimize. Using an ensemble of random circuits to achieve the ``error
per gate'' interpretation as in randomized benchmarking, could be
a waste of resources in this case. One can argue that it is possible
to simply check the expected functionality of the circuit. Yet, even
the case of a single CNOT (or multiple CNOTS) could be quite challenging
since coherent errors are interwind with incoherent errors. Thus,
a deviation from the ideal CNOT map, may not indicate the presence
of an incoherent error. 

\subsection{Locating the error}

Diagnostic processes may have various levels of resolutions. The first
goal is to know if there is an error. The next goal is to locate the
error, and the third is to classify the type of error or ideally provide
the relative weights of various noise mechanisms. Presently, the location
task is carried out by applying holistic methods to smaller systems
(e.g. CNOT's and single-qubit gates) that compose the larger device.
While this method is useful, it has two limitations that one should
keep in mind: i) there could be crosstalk effects that are difficult
to detect when checking a circuits element by element; ii) in a given
circuit and a given initial condition, it could be that a small level
of noise in a good gate is more harmful than other high-noise gates.
One trivial reason could be that this gate is used more time than
the other gates. Thus there is a motivation to locate the error within
the one-circuit diagnostics framework. That is, to run the full circuit
(with many qubits) and mark the hotspots that lead to performance
degradation. Our approach offers a way to locate the noise within
a big circuit without resorting to subsystem benchmarking.

\subsection{Measuring purity}

Purity, entropy, or the R\'{e}nyi entropy are Schur concave functions
that can be used to quantify the amount of randomness (or lack of
it) in quantum or classical systems. The change in the von Neumann
entropy, for example, has a major role in quantum thermodynamics and
quantum information. Purity and R\'{e}nyi entropy have a variety of
applications in quantum information theory as well. Unfortunately,
despite the many insights that these quantities provide, they are
not experimentally friendly. They are nonlinear in the density matrix,
and therefore cannot be directly associated with observables. Rather,
the density matrix should be mapped by measuring a non-scalable number
of observables. The basis in which the density matrix is diagonal,
is a priori unknown, and therefore all elements in the density matrix
have to be evaluated (state tomography).

Several techniques and methods have been suggested to reduce the resources
needed for evaluating the purity. In \cite{Linke2018Renyi} two copies
of the system and a control swap interaction are used for evaluating
the purity. In \cite{ZollerPurityRandomSingleUnitary}, single-qubit
rotations were used to reduce the number of measurements needed for
purity measurement. Reset and reuse of qubits were used in \cite{Subasi2021EntangReset}
to reduce the resources of purity and higher-order R\'{e}nyi entropies
measurements. Finally, matrix product state methods that assume local
buildup of correlation have been studied in \cite{lanyon2017MPStomography}.

Our approach is based on a single observable that is measured at multiple
time points. Thus, the number of observable reduces to one that is
evaluated for each number of cycles. That is, the number of measurements
is equal to the number of cycles which typically ranges in our methods
between three two to five. To avoid overhyping, we stress already
at this point, that our method detects purity changes and not purity.
Furthermore, it is valid only when the change in purity (the dissipation)
is small. 

\subsection{Measuring Entanglement\label{subsec:Measuring-Entanglement intro}}

In the previous section, purity was discussed as an indicator of environment-induced
noise. However, purity can be also used for quantifying entanglement
buildup between two parts of an isolated system. Entanglement is considered
to be the quantum agent responsible for quantum speedups. Nevertheless,
it is quite challenging to measure it, especially if no prior information
on the circuit exists (a ``black box''). If, however, there is an
efficient way of evaluating the local purity of a subsystem, it can
be used to quantify the entanglement of the subsystem to the other
parts of the system. For pure states, a necessary and sufficient condition
for the presence of entanglement is that the purity of a subsystem
is lower than the purity of the total system (which is one for pure
states) \cite{QEntangRMPhorodecki}. Although our results on measuring
purity changes cannot be applied to entanglement measurement as is,
we suggest a modification of the experimental protocol that makes
our methods applicable to entanglement measurements as well.

\section{Our method}

\begin{figure}
\begin{centering}
\includegraphics[width=8.6cm]{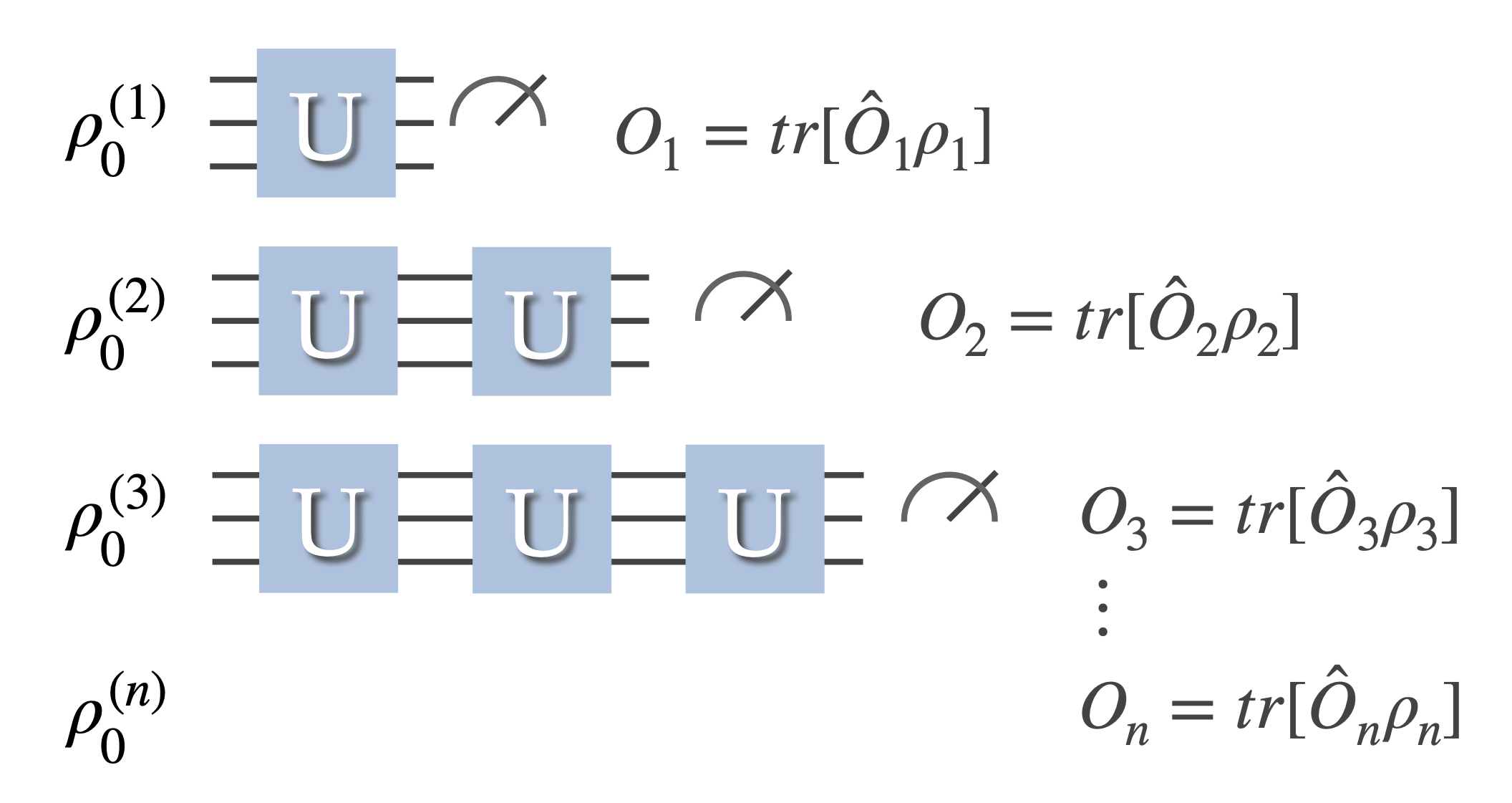}
\par\end{centering}
\caption{\label{fig: scheme}Our methods are based on a set of experiments
with a successively increasing number of repetitions of the basic
quantum circuit (denote by $U$). The measured quantities at the end
of each experiment are combined to form measures that quantify various
dissipation aspects of the system, for example, the total purity change
in the circuit and local dissipation in the qubits. }

\end{figure}

Our method is based on running the circuit of interest multiple times
sequentially, measuring a quantity in each run, and adding the various
results with proper amplitudes. The most general case is illustrated
in Fig. \ref{fig: scheme}. The run we denote by '$k$' is characterized
by having $k$ cycles between the initial state preparation and the
measurement. In the general case, each run may have a different initial
condition $\rho_{0}^{(k)}$ and a different measurement operator $\hat{O}_{k}$
(or a POVM). The expectation value $O_{k}$ is calculated at the end
of each run:
\begin{equation}
O_{k}=\left\langle \hat{O}_{k}\right\rangle =tr[\rho_{k}\hat{O}_{k}].
\end{equation}
For clarity, in each run the circuit is measured many times (``shots'')
to gain sufficiently low $O_{k}$ variance. Next, we compute the following
sum:
\begin{equation}
A_{n}^{O}=\sum_{k=0}^{n}a_{k}^{(n)}O_{k},\label{eq: Agenform}
\end{equation}
and from this sum, we intend to distill information about the purity
change and the hotspots of dissipation. Potentially, in the most general
case, $A_{n}^{O}$ can be a nonlinear function of $\{O_{k}\}_{k=0}^{n}$.

\subsection{The weak action limit of the $\protect\mc S_{n}$ sums}

For measuring the change in purity we set $\rho_{0}^{(k)}=\rho_{0}$
where $\rho_{0}$ can be either a pure state or a mixed state. For
the measurement, we set $\hat{O}_{k}=\rho_{0}$ which lead to the
set observables 
\begin{equation}
R_{k}=tr[\rho_{0}\rho_{k}],
\end{equation}
that we refer to as the survival probability (not to be confused with
the calligraphic $\mc R$ used later on for R\'{e}nyi entropy). For
a pure state $R_{k}$ is the probability to find the system in its
initial state. For mixed states it has the same meaning: writing $\rho_{0}=\sum_{i=0}p_{i}\ketbra ii$
, $R_{k}$ is the probability that the system starts at $\ket 0$
and found to be in $\ket 0$ at the end, plus the probability of starting
at $\ket 1$ and ending in $\ket 1$ and so on. That is, the probability
that a system returns to its initial state without any importance
to what state it was initially in.

We define the coefficients 
\begin{align}
w_{k}^{(n)} & =\frac{2(2n-1)}{n}\frac{n!^{2}}{(n-k)!(n+k)!}(-1)^{k},\\
w_{0}^{(n)} & =2-\frac{1}{n},
\end{align}
and denote the resulting $A_{n}^{\rho_{0}}$ by $\mathcal{S}_{n}$
\begin{equation}
\mathcal{S}_{n}=\sum_{k=0}^{n}w_{k}^{(n)}R_{k},
\end{equation}
where $\mathcal{S}_{1}$ is often too trivial to be of use. The first
few entries are:
\begin{align}
\mathcal{S}_{2} & =\frac{3}{2}R_{0}-2R_{1}+\frac{1}{2}R_{2},\\
\mathcal{S}_{3} & =+\frac{5}{3}R_{0}-\frac{5}{2}R_{1}+R_{2}-\frac{1}{6}R_{3},\\
\mathcal{S}_{4} & =\frac{7}{4}R_{0}-\frac{14}{5}R_{1}+\frac{7}{5}R_{2}-\frac{2}{5}R_{3}+\frac{1}{20}R_{4}.
\end{align}

To understand the physical meaning of these series we move from Hilbert
space to Liouville space. In Liouville space, the density matrix $\rho_{N\times N}$
is flattened into a column ``density vector'' of length $N^{2}$
i.e. $\rho_{N\times N}\to\ket{\rho}_{N^{2}\times1}$. As a result
the Liouville von Neumann equation of motion $id_{t}\rho=[H,\rho]$
becomes:
\begin{align}
id_{t}\ket{\rho} & =H_{L}\ket{\rho}\label{eq: SchrodLio}\\
H_{L} & =H\otimes I_{N}-I_{N}\otimes H^{t}\label{eq: HL}
\end{align}
where $I_{N}$ is the identity operator in the original Hilbert space
($N\times N$) and the subscript $L$ indicates that $H_{L}$ is in
Liouville space. For unitary dynamics $H_{L}$ is hermitian and the
resulting evolution operator in Liouville space is unitary: $\ket{\rho_{t}}=U_{L,t}\ket{\rho_{0}}$
and $U_{L,t}U_{L,t}^{\dagger}=I_{N^{2}}$. If the dynamics is (quantum)
Markovian, the Schr\"{o}dinger-like form (\ref{eq: SchrodLio}) still
holds but now the generator of motion $X=-iH_{L}+L$ does not have
the form (\ref{eq: HL}). In It is easy to obtain L from the Lindblad
equation in Hilbert space (see Sec. \ref{subsec: MatrixElem}) but
we will not need it explicitly here. Due to the Schr\"{o}dinger-like
form, one can write the one cycle evolution operator $\ket{\rho_{k+1}}=U_{L}\ket{\rho_{k}}$
as 
\begin{equation}
U_{L}=e^{-i\mathcal{H}_{L,eff}},
\end{equation}
where $\mathcal{H}_{L,eff}$ is an effective time-independent Hamiltonian
that same evolution as the time-dependent Hamiltonian. Furthermore,
the calligraphic font indicates that we are using for convenience
dimensionless operators that already contain the time duration of
the cycle. In the more general quantum Markovian case we can write
\begin{equation}
K_{L}=e^{-i\mathcal{H}_{L,eff}+\mathcal{L}_{eff}}\doteq e^{x}.\label{eq: Kexp(x)}
\end{equation}

Later on, we shall refer to the operator norm of $\nrm x_{op}$ as
the action of the circuit. Next, we shall employ the fact that the
standard scalar product of matrices in Hilbert space $tr[A^{\dagger}B]$
reads $\braket AB$ in Liouville space where $\bra{\cdot}=\ket{\cdot}^{\dagger}$
as in Hilbert space. Consequently, expectation value can be written
as
\begin{equation}
\left\langle O\right\rangle =tr[O\rho]=\braket O{\rho}.
\end{equation}
Combining this with (\ref{eq: Kexp(x)}) we get $R_{k}=\braOket{\rho_{0}}{e^{kx}}{\rho_{0}}$.
As a result we obtain
\begin{equation}
\mathcal{S}_{n}=\sum_{k=0}^{n}w_{k}^{(n)}\braOket{\rho_{0}}{e^{kx}}{\rho_{0}}=\braOket{\rho_{0}}{\sum_{k=0}^{n}w_{k}^{(n)}e^{kx}}{\rho_{0}}.
\end{equation}
This form shows that that $\mathcal{S}_{n}$ can be seen as an expectation
value of the operator $G_{n}=\sum_{k=0}^{n}w_{k}^{(n)}e^{kx}$. Expressions
similar to $\mathcal{S}_{n}$ have been suggested by the current author
in \cite{UzdinLinke2021PeriodicityIneq}. Yet, these expressions were
used only for constructing inequalities, and their values were not
related to any physical quantities of interest as done here in the
weak action regime. 

Next, we carry out a Markovian open quantum system analysis in the
regime of weak action. We start with an expansion for small $x$:

\begin{align}
\mathcal{S}_{2} & =\left\langle -x+\frac{1}{3}x^{3}+\frac{1}{4}x^{4}+\frac{7}{60}x^{5}+\frac{1}{24}x^{6}+\frac{31}{2520}x^{7}+\frac{1}{320}x^{8}+O\left(x^{9}\right)\right\rangle ,\\
\mathcal{S}_{3} & =\left\langle -x+\frac{1}{6}x^{3}-\frac{11}{120}x^{5}-\frac{1}{12}x^{6}-\frac{239}{5040}x^{7}-\frac{1}{48}x^{8}+O\left(x^{9}\right)\right\rangle ,\\
\mathcal{S}_{4} & =\left\langle -x+\frac{2}{15}x^{3}-\frac{1}{30}x^{5}+\frac{151}{6300}x^{7}+\frac{1}{40}x^{8}+O\left(x^{9}\right)\right\rangle .
\end{align}

Our first key observation is that due to the following property of
Hamiltonians in Liouville space
\begin{equation}
\braOket{\rho}{\mc H_{L}^{2n+1}}{\rho}=0,\label{eq: H2np1eq0}
\end{equation}
all odd terms in $\mathcal{S}_{n}$ drop out when $\mc L=0$. The
proof of (\ref{eq: H2np1eq0}) is given in the Appendix. Thus, when
the evolution is unitary
\begin{align}
\mathcal{S}_{2} & =\frac{1}{4}\left\langle x^{4}\right\rangle +O\left(x^{6}\right),\\
\mathcal{S}_{3} & =-\frac{1}{12}\left\langle x^{6}\right\rangle +O\left(x^{8}\right),\\
\mathcal{S}_{4} & =\frac{1}{40}\left\langle x^{8}\right\rangle +O\left(x^{9}\right).
\end{align}
Consequently, if x is small the $\mc S_{n}$ rapidly converges to
zero as the number of cycles increases. On the other hand if $x=-i\mathcal{H}_{L,eff}+\mathcal{L}_{eff}$
we obtain
\begin{align}
\mathcal{S}_{2} & =-\left\langle \mathcal{L}_{eff}\right\rangle +\frac{1}{3}\left\langle x^{3}\right\rangle +O\left(x^{4}\right),\label{eq: S2}\\
\mathcal{S}_{3} & =-\left\langle \mathcal{L}_{eff}\right\rangle +\frac{1}{6}\left\langle x^{3}\right\rangle +O\left(x^{5}\right),\\
\mathcal{S}_{4} & =-\left\langle \mathcal{L}_{eff}\right\rangle +\frac{2}{15}\left\langle x^{3}\right\rangle +O\left(x^{5}\right).\label{eq: S4}
\end{align}
Note, that this time $\left\langle x^{3}\right\rangle \neq0$ since
it contains term like $H^{2}L$ and $L^{3}$. Nevertheless, we will
assume that in the weak action regime, $x^{3}$ has a negligible contribution
to $\mathcal{S}_{2}$. Furthermore, by taking $2\mathcal{S}_{3}-\mathcal{S}_{2}$
we get
\begin{equation}
2\mathcal{S}_{3}-\mathcal{S}_{2}=-\left\langle \mathcal{L}_{eff}\right\rangle +O\left(x^{4}\right).
\end{equation}
By increasing the cycle number it is possible to eliminate higher-order
corrections and make the evaluation $\left\langle \mathcal{L}_{eff}\right\rangle $
applicable to circuits with larger action. This, however, cannot be
done indefinitely, since the measurement uncertainty tends to increase
when adding more cycles to eliminate higher-order corrections. In
the rest of the paper, for brevity, we drop the ``eff'' and ``L''
subscripts from the generators of motion.

\section{Applications}

\subsection{Measuring the matrix element of the dissipator\label{subsec: MatrixElem}}

Equations (\ref{eq: S2})-(\ref{eq: S4}) show that the $\mathcal{S}_{n}$'s
provide a direct information on the ``dissipator matrix element''
in Liouville space. This allows to learn about the active noise mechanisms
in a given circuit and also use this information to predict how they
affect other circuits. As a first example let us look at a single
spin spontaneous emission. The annihilation operator in Hilbert space
is $a=\left(\begin{array}{cc}
0 & 0\\
1 & 0
\end{array}\right)$ and the corresponding Liouvillian is:
\begin{equation}
L_{spon}=\xi[a\otimes(a^{\dagger})^{t}-\frac{1}{2}a^{\dagger}a\otimes I_{2}-\frac{1}{2}I_{2}\otimes a^{\dagger}a],
\end{equation}
where $\xi$ corresponds to the decay rate (the decay time ``$T_{1}$''
is equal to $1/\xi$). Using $\ket{\uparrow_{L}}=\{1,0,0,0\}$, $\ket{\uparrow_{L}}=\{0,0,0,1\}$
and $\ket{+_{L}}=\frac{1}{2}\{1,1,1,1\}$ to denote up, down and plus
states in Liouville space, we find that 
\begin{align}
\braOket{\uparrow_{L}}{L_{spon}}{\uparrow_{L}} & =-\xi,\\
\braOket{\downarrow_{L}}{L_{spon}}{\downarrow_{L}} & =0,\\
\braOket{+_{L}}{L_{spon}}{+_{L}} & =-\frac{1}{4}\xi.
\end{align}

A depolarizing channel can be written as $L_{depol}=L_{spon}+L_{spon}(a\leftrightarrow a^{\dagger})$
and it satisfies:
\begin{align}
\braOket{\uparrow_{L}}{L_{depol}}{\uparrow_{L}} & =-\xi,\\
\braOket{\downarrow_{L}}{L_{depol}}{\downarrow_{L}} & =-\xi,\\
\braOket{+_{L}}{L_{depol}}{+_{L}} & =-\frac{1}{2}\xi.
\end{align}

For a pure decoherence channel the Lindblad operator is the Pauli
$\sigma_{z}$ and the resulting Liouvillian is:
\begin{equation}
L_{decoh}=\frac{1}{2}\xi[\sigma_{z}\otimes\sigma_{z}-I_{2}\otimes I_{2}],
\end{equation}
and we get

\begin{align}
\braOket{\uparrow_{L}}{L_{decoh}}{\uparrow_{L}} & =0,\\
\braOket{\downarrow_{L}}{L_{spon}}{\downarrow_{L}} & =0,\\
\braOket{+_{L}}{L_{spon}}{+_{L}} & =-\frac{1}{2}\xi.
\end{align}
Thus by evaluating $-\mathcal{S}_{2}$ (or $\mathcal{S}_{n>2}$) for
the initial states $\ketbra{\downarrow}{\downarrow}$, $\ketbra{\uparrow}{\uparrow}$
and $\ketbra ++$ it is possible to identify the decay mechanism of
the spin. If the channel is thermal, one can use the ratio $\braOket{\uparrow_{L}}{L_{\beta}}{\uparrow_{L}}/\braOket{\downarrow_{L}}{L_{\beta}}{\downarrow_{L}}=e^{-\beta\omega}$
for evaluating the inverse temperature $\beta$ given the energy gap
of the qubit $\omega$. We conclude that by using different initial
condition it is possible to investigate the nature of some unknown
environment.

The reader may be puzzled at this point since the decoherence time,
for example, could be simply evaluated by measuring $\sqrt{\left\langle \sigma_{x}\right\rangle ^{2}+\left\langle \sigma_{y}\right\rangle ^{2}}$
as a function of time in the absence of driving. Yet, our method enables
to extract information on the dissipator under arbitrary weak driving
which fits the spirit of ``one-circuit diagnostics'' rather than
using a dedicated circuit for the job. 

In the presence of multiple qubits evaluation of $\mathcal{S}_{2}$
for the initial state $\ket +\otimes\ket +\otimes\ket +...$ will
yield the sum of the local decoherence rates which represent the leading
term in the purity loss in the total circuit as explained in the next
section. Interestingly, even when running the whole circuit it is
possible to evaluate the individual qubit dissipation rate using the
methods we present in Sec. \ref{subsec: hot-spots} and in Fig. \ref{fig: Location}.

\subsection{Measuring small purity changes using $\mathcal{S}_{n}$\label{subsec:Measuring-small-purity}}

Let us look at the change in purity after one cycle in the weak action
regime:
\begin{align}
\Delta tr\rho^{2} & =\braOket 0{e^{x^{\dagger}}e^{x}}0-\braket 00\nonumber \\
 & =\braOket 0{(1+x^{\dagger}+x^{\dagger2}/2)(1+x+x^{2}/2)}0-\braket 00+O(x^{3},x^{2}x^{\dagger},..)\nonumber \\
 & =\left\langle x+x^{\dagger}+x^{\dagger2}/2+x^{2}/2+x^{\dagger}x\right\rangle +O(x^{3},x^{2}x^{\dagger},..)\nonumber \\
 & =\left\langle 2\mc L-i\{\mc H,\mc L^{\dagger}\}+i\{\mc H,\mc L\}+\mc L^{\dagger2}/2+\mc L^{2}/2+\mc L^{\dagger}\mc L\right\rangle +O(x^{3},x^{2}x^{\dagger},..).\label{eq: purity small H}
\end{align}

Thus when $\mc L$ and $\mc H$ are small the second-order terms are
negligible and we get
\begin{equation}
\Delta tr\rho^{2}=-2S_{n}+O(\mc L^{2},\mc L\mc H,\mc H\mc L).
\end{equation}

\begin{figure}
\begin{centering}
\includegraphics[width=8.6cm]{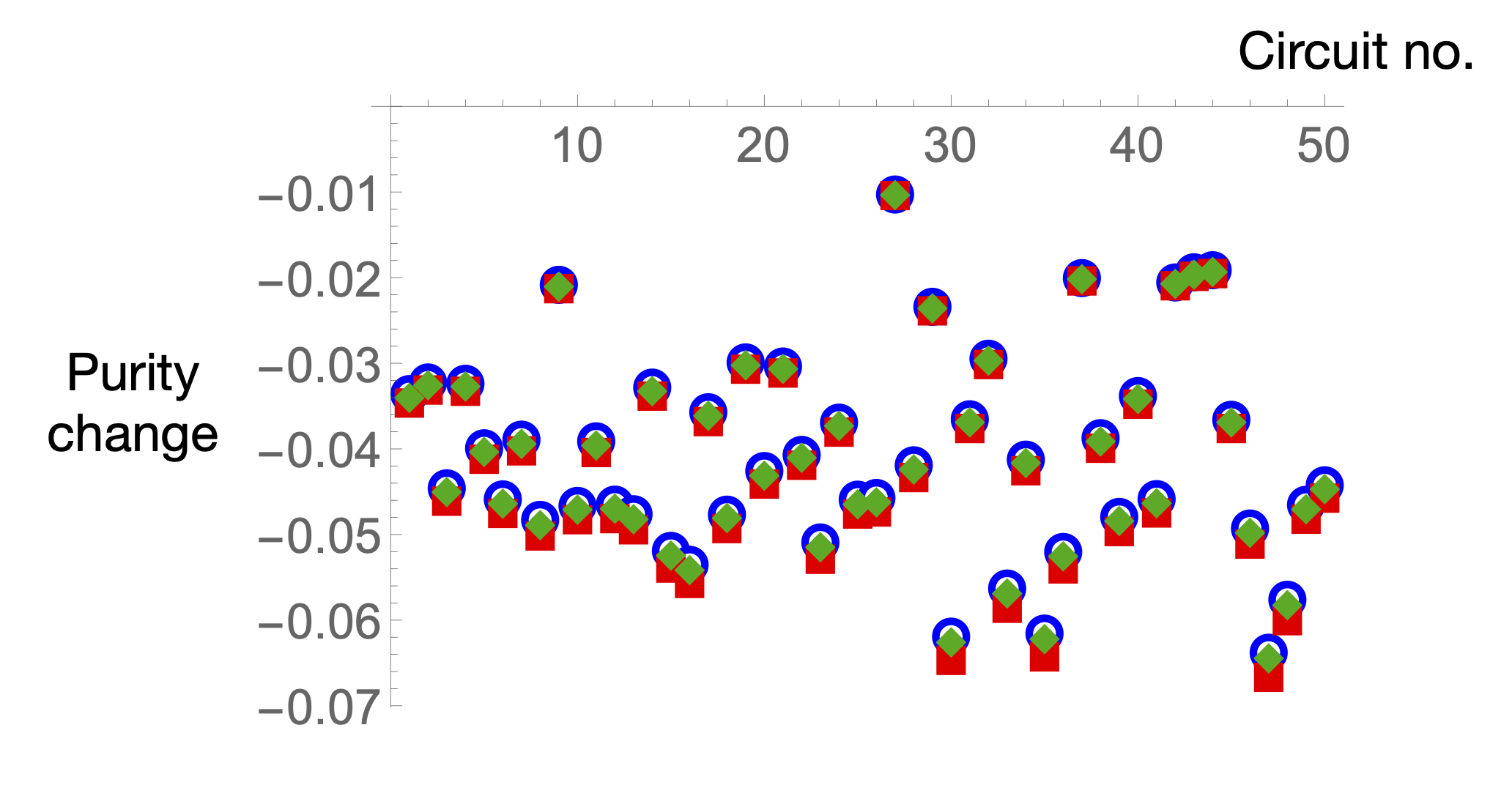}
\par\end{centering}
\caption{\label{fig: Purity}Estimating the purity change in a four-qubit weak
action random circuits with random decoherence in each qubit. (b)
Each point on the horizontal axis corresponds to a different circuit
and different decoherence rates. The blue circles mark the exact purity
change, the red square corresponds to our $-2\mathcal{S}_{2}$ method,
and green diamonds stand for the refinement $-2\mathcal{S}_{2}+2\mathcal{S}_{2}^{2}$,
in Eq. (\ref{eq: L2 correction}).}

\end{figure}

For some dissipators, the expression (\ref{eq: purity small H}) simplifies
and further approximations can be made. For example, for depolarizing
channel and for all Hermitian Lindblad operators, e.g., decoherence
operators, it holds that $\mc L=\mc L^{\dagger}$  and therefore:
\begin{equation}
\Delta tr\rho^{2}=2\left\langle \mc L\right\rangle +2\left\langle \mc L^{\dagger}\mc L\right\rangle +O(x^{3},x^{2}x^{\dagger},..).
\end{equation}
Since $\left\langle \mc L^{\dagger}\mc L\right\rangle \ge\left\langle \mc L\right\rangle ^{2}$
it follows that $\Delta tr\rho^{2}\ge\left\langle \mc L\right\rangle +2\left\langle \mc L\right\rangle ^{2}$.
For pure state $\left\langle \mc L\right\rangle \le0$ so $2\left\langle \mc L\right\rangle $
slightly overestimates the rate and the term $2\left\langle L\right\rangle ^{2}$
reduces this overestimation. Hence we can write
\begin{equation}
\Delta tr\rho^{2}\simeq2\left\langle \mc L\right\rangle +2\left\langle \mc L\right\rangle ^{2}=-2\mathcal{S}_{2}+2\mathcal{S}_{2}^{2}.\label{eq: L2 correction}
\end{equation}
Figure \ref{fig: Purity} test our purity change estimation on random
circuits. Each point on the horizontal axis corresponds to a different
circuit. A randomly chosen (weak) Hamiltonian generates the unitary
drive and each qubit undergoes decoherence at a different random rate
(the rate change from one circuit to another but not during the evolution).
The Blue circles correspond to the exact value of the purity change
in one cycle $tr\rho_{1}^{2}-tr\rho_{0}^{2}$, and the red squares
correspond to our $-2\mathcal{S}_{2}$ estimation. Since the Lindblad
operator is hermitian (decoherence) we also plot in green diamonds
the refinement (\ref{eq: L2 correction}) which substantially improves
the purity estimation accuracy for higher purity changes.

\subsection{Measuring entanglement}

In continuation to Sec. (\ref{subsec:Measuring-Entanglement intro}),
one of the entanglement measures for a pure state $\rho_{AB}$ in
a bipartite system $AB$ is directly related to the purity, or more
accurately to the R\'{e}nyi entanglement entropy of order two \cite{QEntangRMPhorodecki}:
\begin{equation}
\mc R_{2}=-\ln tr[\rho_{A}^{2}],
\end{equation}
where $\rho_{A}=tr_{B}\rho_{AB}$. Thus, methods for evaluating the
purity of a subsystem can be exploited to quantify entanglement as
well. While in some methods for evaluating the purity, the application
to entanglement measurement is straightforward, here there is an interesting
difference that requires a modification in the $\mc S_{n}$ experimental
protocol. 

\begin{figure}
\begin{centering}
\includegraphics[width=8.6cm]{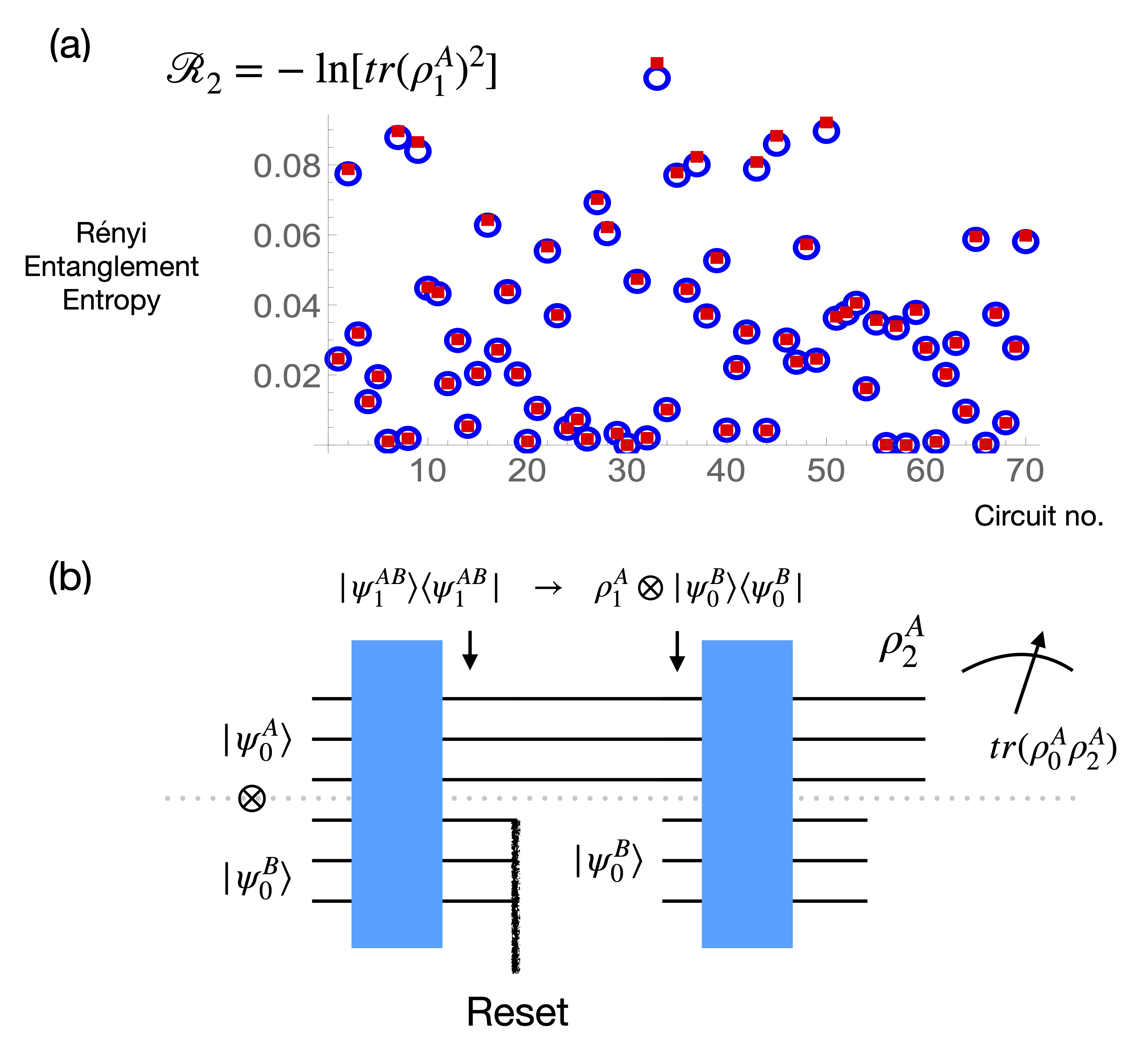}
\par\end{centering}
\caption{\label{fig: Entang}(a) A numerical simulation of measuring the entanglement
in a six-qubit system using our periodic reset protocol (b). The blue
circles represent the exact value of the R\'{e}nyi 2 entanglement
entropy between two sets of three qubits. The red squares correspond
to our $\mathcal{S}_{2}^{reset}$ method. Each point on the horizontal
axis corresponds to a different small action random circuit. As expected,
the method works well when the entanglement is small (roughly $0.07$
in this case). }

\end{figure}

The $\mc S_{n}$ method for evaluating the purity change in the first
cycle (Sec. \ref{subsec:Measuring-small-purity}) assumes that the
dynamics is periodic and Markovian, or alternatively stated, a periodic
CP map. However, if the AB system evolves unitarily, the reduced dynamics
of $\rho_{A}$ is typically non-markovian and therefore $\mc S_{n}$
will not yield information on the purity change in A. To induce a
periodic CP map we use the following recipe. Assume the initial state
is a product pure product state
\begin{equation}
\rho_{0}=\ketbra{\psi_{A}}{\psi_{A}}\otimes\ketbra{\psi_{B}}{\psi_{B}}.
\end{equation}
After applying one cycle of the unitary evolution we reset subsystem
B to its initial state and only then perform the unitary evolution
of the next cycle. In the general case, the same is carried out for
the other cycles so that:
\begin{align}
\rho_{n} & =\rho_{n,A}\otimes\ketbra{\psi_{B}}{\psi_{B}},\\
\rho_{n+1} & =U(\rho_{n,A}\otimes\ketbra{\psi_{B}}{\psi_{B}})U^{\dagger}.
\end{align}
The resetting generate a periodic CP map for which we can apply our
method. The resetting changes evolution, however for the first cycle
$\rho_{A,1}^{\text{no reset}}=\rho_{A,1}^{\text{reset}}$. Thus, by
evaluating the purity change in the first cycle of the system with
the reset, we obtain the purity change in the original system. Starting
in a pure state we finally get
\begin{equation}
\Delta\mc R_{2}=\mc R_{2}^{fin}=-\ln tr\rho_{fin,A}^{2}=-\ln(1+\Delta tr\rho_{A}^{2})\simeq\mathcal{S}_{n,A}^{reset},\label{eq: SnReset}
\end{equation}
where $\mathcal{S}_{n,A}^{reset}$ stands for the purity change evaluation
in subsystem A with the reset protocol. In Fig. \ref{fig: Entang}
we took a six-qubit system and partitioned it into two parts with
three qubits each. Random Hamiltonians in the space of the six qubits
were used to weakly entangle the two parts. To make sure the dynamics
is in the weak action regime the operator norm of the Hamiltonians
was restricted to values below 0.8. The blue circles in Fig. \ref{fig: Entang}
show the exact calculation of $\mc R_{2}$, and the red square corresponds
to our reset method (\ref{eq: SnReset}). Fig. \ref{fig: Entang}b
illustrates the two-cycle run in the reset protocol.

\subsection{Dissipation hot spots - locating the incoherent errors in the circuit\label{subsec: hot-spots}}

In trying to apply our method to other quantities (observables) its
useful to point the key elements in the $\mathcal{S}_{n}$ method:
\begin{enumerate}
\item The choice of $w_{k}^{(n)}$ lead to the cancellation of some of the
leading even-order terms in the weak action regime.
\item The odd terms are zero when $x=i\mc H$, $\braOket{\rho_{0}}{\mc H^{2k+1}}{\rho_{0}}=0$.
\item For periodic CP maps, the first-order term is not zero, but it depends
only on the dissipative part of the dynamics. Thus the first-order
term can be used to study the interaction of the system with the environment.
\end{enumerate}
Property \#1 makes no use of the fact that the quantity of interest
is purity. Property \#2 was proved in the first part of the Appendix
for survival probabilities. That is, the observable $A$ was equal
to the initial density matrix $\rho_{0}$. As it turn out, for a general
observable $A=A^{\dagger}$ in Hilbert space, $\braOket A{\mc H_{L}^{2n+1}}{\rho_{0}}\neq0$.
Yet, we show in the second part of the appendix that for observables
satisfying $[A,\rho_{0}]=0$ it holds that
\begin{equation}
\braOket A{\mc H_{L}}{\rho_{0}}=0.\label{eq: AHeq0}
\end{equation}
Next, we set $\text{\ensuremath{\hat{O}_{k}}}=A$ in (\ref{eq: Agenform})
and for the coefficients we choose $a_{k}^{(n)}=w_{k}^{(n)}$ as before,
and we get 
\begin{equation}
\mathcal{S}_{n}^{A}=\sum_{k=0}^{n}w_{k}^{(n)}A_{k}=\sum_{k=0}^{n}w_{k}^{(n)}tr[A\rho_{k}],
\end{equation}
that, as before, we can write as
\begin{equation}
\mathcal{S}_{n}^{A}=\sum_{k=0}^{n}a_{k}^{(n)}\braOket A{e^{kx}}{\rho_{0}}=\braOket A{\sum_{k=0}^{n}a_{k}^{(n)}e^{kx}}{\rho_{0}}.
\end{equation}
Using (\ref{eq: AHeq0}) we obtain

\begin{align}
\mathcal{S}_{2}^{A} & =-\braOket A{\mc L}{\rho_{0}}+O\left(x^{3}\right),\\
2\mathcal{S}_{3}^{A}-\mathcal{S}_{2}^{A} & =-\braOket A{\mc L}{\rho_{0}}+O\left(x^{4}\right).\label{eq: A34}
\end{align}

Note, that here the third order does not cancel out. Yet, by running
one more cycle and using (\ref{eq: A34}) the third order can be eliminated
for (for unitary and nonunitary evolution both). To understand what
this quantity means we evaluate the ``dissipative change'' in $\left\langle A\right\rangle $:

\begin{align}
\Delta_{\text{diss}}\left\langle A\right\rangle  & =\Delta\left\langle A\right\rangle -\Delta_{\text{no diss}}\left\langle A\right\rangle \nonumber \\
 & =\braOket A{e^{x}}{\rho_{0}}-\braket A{\rho_{0}}-(\braOket A{e^{-i\mc H}}{\rho_{0}}-\braket A{\rho_{0}})\nonumber \\
 & =\braOket A{(1+x+x^{2}/2}{\rho_{0}}-\braOket A{(1-i\mc H-\mc H^{2}/2}{\rho_{0}}+O(x^{3},x^{2}x^{\dagger},..)\nonumber \\
 & =\bra A\mc L+(i\mc H+\mc L)^{2}/2+\mc H{}^{2}/2\ket{\rho_{0}}+O(x^{3},x^{2}x^{\dagger},..)\nonumber \\
 & =\bra A\mc L+\mc L^{2}/2+i\{\mc L,\mc H\}/2\ket{\rho_{0}}+O(x^{3},x^{2}x^{\dagger},..).
\end{align}

If the Lindblad operators that generates $\mc L$ are hermitian and
the dissipator is incoherent i.e. $\bra{\text{cohences}}\mc L\ket{\text{diagonal state}}=0$
(the map does not create coherences when starting in a diagonal state)
then $\braOket A{\{\mc L,\mc H\}}{\rho_{0}}=0$. Another option for
eliminating the $\mc{HL}$ term is to run another circuit with $\mc H\to-\mc H$,
and measure the mean $\frac{1}{2}[\mathcal{S}_{2}^{A}(\mc H)+\mathcal{S}_{2}^{A}(-\mc H)]$.
Assuming that $\{\mc L,\mc H\}$ is either zero or removed we get:
\begin{align}
\text{\ensuremath{\Delta_{diss}\left\langle A\right\rangle }} & =-\mathcal{S}_{2}^{A}+O(\mc L^{2})+O(x^{3}),\\
\text{\ensuremath{\Delta_{diss}\left\langle A\right\rangle }} & =-2\mathcal{S}_{3}^{A}+\mathcal{S}_{2}^{A}+O(\mc L^{2})+O(x^{4}),\label{eq: DAdiss34}\\
\text{\ensuremath{\Delta_{diss}\left\langle A\right\rangle }} & =-5\mathcal{S}_{4}^{A}+4\mathcal{S}_{3}^{A}+O(\mc L^{2})+O(x^{5}).\label{eq: x5}
\end{align}
Note that the driving term can be significant so the cancellation
of the third order can be important.

\begin{figure}
\begin{centering}
\includegraphics[width=9cm]{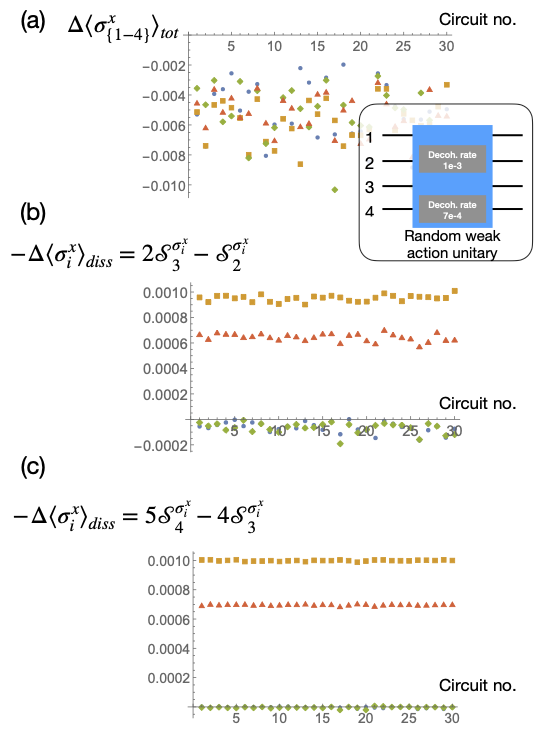}
\par\end{centering}
\caption{\label{fig: Location} In this example, four qubits interact via a
random Hamiltonian (inset). Qubit \#2 undergoes decoherence at a rate
$\xi_{2}=0.001$ and the decoherence rate of qubit \#4 is $\xi_{4}=0.0007$.
Qubit \#1 and \#3 are not directly dissipated. In all plots, the x
axis corresponds to different random Hamiltonians. Figure (a) shows
that the change in the $\sigma^{x}$ expectation values of the qubits
is roughly the same for all qubits. Thus it is not possible to tell
which qubits are losing coherence. However, when using our method
{[}Fig. (b) \& (c){]} to evaluate the dissipative contribution to
the change in $\left\langle \sigma_{i}^{x}\right\rangle $, it becomes
clear that qubit \#2 (orange squares) and qubit \#4 (red triangles)
have clear dissipative change while qubit \#1 (blue circles) and qubit
\#3 (green diamonds) experience no dissipative change. In Fig. (b)
three cycles are used, which lead to fourth-order correction in the
action (\ref{eq: DAdiss34}). As a result, the dissipative change
of qubits 1\# and 3\# is not exactly zero. In (c) we use 4 cycles,
and consequently, the correction is of order five in the action (\ref{eq: x5}).
Here it is clear that our method correctly retrieves the values of
the decoherence rate $\xi_{2},\xi_{4}$ and the null rates of qubits
1 and 3. }
\end{figure}

To illustrate our finding, in Fig. \ref{fig: Location}, we consider
a four-qubit example with a random weak unitary driving. Qubits 2
and 4 are noisy (inset). The driving Hamiltonian is chosen with random
elements in the interval $\left\{ \pm0.1\pm0.1i\right\} $ and the
time interval is $T=1$ which leads to an average action of $[max(H)-min(H)]T=0.08$.
The initial state is $\ket{++++}$. The decoherence rate of qubit
\#2 is $\xi_{2}=1e-3$ ($\tau_{(2)}=1000$) and in qubit \#$4$ it
is $\xi_{4}=7e-4$ ($\tau_{(4)}=1428.6$). Without driving the coherence
$\left\langle \sigma_{x}\right\rangle $ of qubit 2 decays from $1$
to $0.9995$. Qubits 1 and 3 are not directly dissipated, they only
interact with dissipative qubits. The observables of interest are
the $\left\langle \sigma^{x}\right\rangle $ of each qubit. Before
studying the dissipative change in $\left\langle \sigma^{x}\right\rangle $
of each qubit we plot in Fig. \ref{fig: Location}a the ``total''
change (driving + dissipation) in qubit $i$, $\Delta\left\langle \sigma_{i}^{x}\right\rangle =tr[(\rho_{1}-\rho_{0})\sigma_{i}^{x}]$.
The dissipative qubits 2 \& 4 seem no different from the non-dissipative
qubits 1 \& 2. Yet, by running three {[}Fig. \ref{fig: Location}(b){]}
or four {[}Fig. \ref{fig: Location}(c){]} cycles, we can employ our
methods and correctly evaluate the decoherence rate of the different
qubits despite the random unitary in the background.

Finally, let us point out that the ``hotspot'' may not be individual
qubits but gates. If the gate mechanism itself involves some interaction
with the environment, then only when this gate is activated noise
will appear in the system. The qubits that are affected by this gate
will appear as noisy.

\subsubsection*{}

\section{Relieving the weak action restriction}

Due to the weak action validity regime, it is not possible to immediately
apply the methods here presented to any quantum circuit. Yet there
are two alternatives that enable indirectly to address any circuit
provide the noise is sufficiently small. 

\subsection{``Weakened circuits''}

The first option is to have a ``weak version'' of the original circuit.
This can be done by simply making all the RF/laser control weaker
by a factor $\gamma<1$. Assuming the dissipation in the original
circuit is weak the new evolution operator is $e^{-i\gamma\mc H_{L}^{eff}+\mc L^{eff}}$
will be in the weak action regime and our methods can be applied.
Although we presently do not know how to analytically connect the
change in purity in the weakened circuit to the purity change in the
original large action circuit, one can argue that the weak action
version has the same implementation map and it is susceptible to the
same noise mechanism as the original circuit. In particular, if non-negligible
dissipation effects appear already at the weakened version they are
unlikely to just fade away in the original circuit. Thus it could
be a good practice to first use the weakened version to optimize the
performance of the device and only then proceed to check the original
circuit. 

Another way of using the weakened circuit is to moderately increase
the action of the weakened circuit but use more cycles to evaluate
higher-order $\mc S_{n}$'s {[}and combinations of $\mc S_{n}$'s
as in (\ref{eq: x5}){]}. Interestingly, when using sufficiently weak
local dissipators on each qubit, we find that the purity loss is roughly
independent of the strength of the drive (the action of the noiseless
circuit) even if it is very strong. We observed this behavior when
the purity loss was $\sim0.03$ or less. If this finding is general,
it paves the way to using weakened circuits for quantifying the noise
in arbitrary circuits that are subjected only to local decay and decoherence.

\subsection{Using the inverse circuit}

Another alternative is to implement the circuit and immediately after
implementing the inverse circuit. Since the dissipation mechanisms
are the same for the inverse circuit, it follows that if the evolution
operator (with the dissipation) in Liouville space is given by $K=e^{-i\mc H+\mc L}$,
then the evolution operator of the inverse circuit is given $K_{I}=e^{-i\mathcal{H}_{I}+\mc L}$
where $\mathcal{H}_{I}=-\mathcal{H}+d\mathcal{H}$ where $d\mathcal{H}$
represent potential coherent errors. For simplicity we Next we compare
the purity in one cycle:
\begin{equation}
\Delta tr[\rho^{2}]_{K}=\braOket{r_{0}}{K^{\dagger}K}{r_{0}}-\braket{r_{0}}{r_{0}}
\end{equation}
to the purity create by the circuit and its noisy inverse
\begin{equation}
\Delta tr[\rho^{2}]_{K_{I}K}=\braOket{r_{0}}{(K_{I}K)^{\dagger}K_{I}K}{r_{0}}-\braket{r_{0}}{r_{0}}
\end{equation}

Treating $L$ as small and using the derivative of the exponential
map we get:
\begin{equation}
K=e^{-i\mc H+\mc L}=e^{-i\mc H}\{1+\sum_{k=0}^{\infty}\frac{1}{k+1}[(+i\mc H)^{(k)},\mc L]\}+O(\mc L^{2})\}
\end{equation}
where $[A^{\ensuremath{(1)}},B]=[A,B]$, $[A^{\ensuremath{(2)}},B]=[A,[A,B]]$,
$[A^{\ensuremath{(3)}},B]=[A,[A,[A,B]]]$ and so on. After some algebra
we show in Appendix II that: 
\begin{equation}
\Delta tr[\rho^{2}]_{K}=\braOket{r_{0}}{1+\sum_{k=0}^{\infty}\frac{2}{k+1}[(+i\mc H)^{(k)},\mc L]}{r_{0}}+O(\mc L^{2})
\end{equation}
which is an extension of eq. (\ref{eq: purity small H}) to large
$\mc H$. Next, we calculate the change in purity of the inverse circuit.
In Appendix II we get:

\begin{equation}
K_{I}K=1+\sum_{k=0}^{\infty}\frac{2}{k+1}[(+i\mc H)^{(k)},\mc L]+O(\mc L^{2})
\end{equation}
and finally the purity 
\begin{equation}
(K_{I}K)^{\dagger}K_{I}K=1+\sum_{k=0}^{\infty}\frac{2}{k+1}\{[(-i\mc H)^{(k)},\mc L]+[(-i\mc H)^{(k)},\mc L]^{\dagger}\}+O(\mc L^{2})
\end{equation}
this is a positive operator so so for a state $\phi$ it holds that
$\braOket{\phi}{[(-i\mc H)^{(k)},\mc L]^{\dagger}}{\phi}=\braOket{\phi}{[(-i\mc H)^{(k)},\mc L]}{\phi}$
as a result we get that:
\begin{equation}
\braOket{r_{0}}{(K_{I}K)^{\dagger}K_{I}K}{r_{0}}=1+\sum_{k=0}^{\infty}\frac{4}{k+1}[(-i\mc H)^{(k)},\mc L]
\end{equation}
and we get 
\begin{align}
\Delta tr[\rho^{2}]_{K_{I}K} & =\braOket{r_{0}}{\sum_{k=0}^{\infty}\frac{4}{k+1}[(-i\mc H)^{(k)},\mc L]}{r_{0}}+O(\mc L^{2})\nonumber \\
 & =2\braOket{r_{0}}{\sum_{k=0}^{\infty}\frac{2}{k+1}[(-i\mc H)^{(k)},\mc L]}{r_{0}}+O(\mc L^{2})\nonumber \\
 & =2\Delta tr[\rho^{2}]_{K}+O(\mc L^{2})
\end{align}

$\delta H$ represents the presence a potential unitary error. We
now make the approximation $\Delta tr[\rho^{2}]_{\delta\mc H\neq0}=\Delta tr[\rho^{2}]_{\delta\mc H=0}$
which means that a slight coherent error will not affect the leading
orders in the purity loss of the circuit+inverse system. Thus we can
replace $K'^{\dagger}$ with $K^{\dagger}$. Next, we study the purity
change in the circuit that contains the original circuit and its inverse.
To make a point we start with $\rho_{0}$ which can be either mixed
or pure. 

\section*{Concluding remarks}

In this paper, we derived several tools that are based on collecting
data from several different repetitions of the same circuit to retrieve
information on the dissipative effects that take place in the quantum
device (circuit). It was shown how to measure the purity change and
even how to differentiate between different noise mechanisms. Our
approach also enables to locate the error in specific parts of the
circuit while running the whole circuit. Thus, it saves the need to
evaluate each part of the system separately in order to isolate the
problem. Since we use only a few observable regardless of the system
size, our approach is scalable. Another interesting application that
is unrelated to diagnostics, is entanglement measurement in isolated
systems. We demonstrated that by incorporating a reset to part of
the system in our basing protocol it is possible to measure the R\'{e}nyi
$2$ entanglement. While these methods can be valuable for developers,
they can also be useful for end-users that want to verify the performance
of the device just before using it. 

\section*{Appendix I - Derivation of $\protect\braOket{\rho_{0}}{H_{L}^{2n+1}}{\rho_{0}}=0$}

Let start with simple case of n=0. By definition H it holds that 
\begin{align}
\braOket{\rho_{0}}{H_{L}}{\rho_{0}} & =\braket{\rho_{0}}{[H,\rho_{0}]}=tr(\rho_{0}^{\dagger}[H,\rho_{0}])\nonumber \\
 & =tr(\rho_{0}H\rho_{0}-\rho_{0}\rho_{0}H)=0\label{eq: rHLreq0ap}
\end{align}
note that the only needed property of property $\rho_{0}$ is hermiticity.
\begin{align}
\braOket{\rho_{0}}{H_{L}^{2n+1}}{\rho_{0}} & =-(\braOket{\rho_{0}}{H_{L}^{n})H_{L}(H_{L}^{n}}{\rho_{0}})\nonumber \\
 & =\braOket r{H_{L}}r=tr(r^{\dagger}[H,r])\nonumber \\
 & =tr(r^{\dagger}Hr-r^{\dagger}rH)=tr\{(rr^{\dagger}-r^{\dagger}r)H\}.
\end{align}
Thus if $r^{\dagger}=r$ (in Hilbert space)

\begin{equation}
[H,[H,[H,\rho_{0}]]]^{\dagger}=[[[\rho_{0},H],H],H]=(-1)^{n}[H,[H,[H,\rho_{0}]]].
\end{equation}

In the second part of the appendix show that if and observable $A=A^{\dagger}$
satisfies $[A,\rho_{0}]=0$ it holds that
\begin{equation}
\braOket A{H_{L}}{\rho_{0}}=0.
\end{equation}
The derivation is straight forward and similar to (\ref{eq: rHLreq0ap})
\begin{align}
\braOket A{H_{L}}{\rho_{0}} & =\braket{A^{\dagger}}{[H,\rho_{0}]}\nonumber \\
 & =tr(A^{\dagger}H\rho_{0}-A^{\dagger}\rho_{0}H])\nonumber \\
 & =tr([\rho_{0},A^{\dagger}]H)=0.
\end{align}

\section*{Appendix II - purity change with the inverse circuit}

Using the identity

\begin{equation}
[A^{(k)},B]^{\dagger}=(-1)^{k}[(A^{\dagger})^{(k)},B^{\dagger}]=[(-A^{\dagger})^{(k)},B^{\dagger}]\label{eq: dagger ident},
\end{equation}
we get $[(i\mc H)^{(k)},\mc L]^{\dagger}=[(i\mc H)^{(k)},\mc L^{\dagger}]$
so that
\begin{align}
K^{\dagger}K & =\{1+\sum_{k=1}^{\infty}\frac{1}{k+1}[(i\mc H)^{(k)},\mc L]^{\dagger}\}+O(\mc L^{\dagger2})\}e^{+i\mc H}\nonumber \\
 & \times e^{-i\mc H}\{1+\sum_{k=1}^{\infty}\frac{1}{k+1}[(i\mc H)^{(k)},\mc L]\}+O(\mc L^{2})\}.
\end{align}
Using the fact $\braOket{\phi}{[(-i\mc H)^{(k)},\mc L]}{\phi}$ is
real it holds that 
\begin{equation}
\braOket{\phi}{[(i\mc H)^{(k)},\mc L]}{\phi}=\braOket{\phi}{[(i\mc H)^{(k)},\mc L]^{\dagger}}{\phi},
\end{equation}

and keeping only $O(\mc L)$ we get:
\begin{equation}
K^{\dagger}K=1+(\sum_{k=1}^{\infty}\frac{2}{k+1}[(i\mc H)^{(k)},\mc L]+h.c.)+O(\mc L^{2}).
\end{equation}

\subsection*{Dealing with the inverse circuit}

To deal with $K_{I}$ we write it as $K_{I}=(e^{-i\mathcal{H}+\mathcal{L}^{\dagger}})^{\dagger}$
\begin{equation}
e^{-i\mathcal{H}+\mathcal{L}^{\dagger}}=e^{-i\mc H}\{1+\sum_{k=0}^{\infty}\frac{1}{k+1}[(+i\mc H)^{(k)},\mc L^{\dagger}]\}+O(\mc L^{\dagger2})\}.
\end{equation}
Thus 
\begin{align}
K_{I}K & =\{1+\sum_{k=1}^{\infty}\frac{1}{k+1}[(+i\mc H)^{(k)},\mc L^{\dagger}]^{\dagger}\}+O(\mc L^{2})\}\nonumber \\
 & \times\{1+\sum_{k=1}^{\infty}\frac{1}{k+1}[(+i\mc H)^{(k)},\mc L]\}+O(\mc L^{2})\}=\\
 & 1+\sum_{k=1}^{\infty}\frac{2}{k+1}[(+i\mc H)^{(k)},\mc L]\}+O(\mc L^{2}).
\end{align}
where we have used (\ref{eq: dagger ident}) to get

\begin{equation}
\sum_{k=0}^{\infty}\frac{1}{k+1}[(i\mc H)^{(k)},\mc L^{\dagger}]^{\dagger}=\sum_{k=0}^{\infty}\frac{1}{k+1}[-[(-i\mc H)^{\dagger}]^{(k)},\mc L].
\end{equation}

\bibliographystyle{unsrt}
\bibliography{RaamCite1}

\end{document}